\documentclass[journal]{IEEEtran}
\usepackage{graphicx}
\usepackage[normalem]{ulem}
\usepackage{amssymb}
\ifCLASSINFOpdf

\else

\fi

\begin{document}

\title{Multi-user quantum key distribution with collective eavesdropping detection over collective-noise channels}

\author{Wei~Huang, Qiao-Yan~Wen, Bin~Liu, Fei~Gao 
\thanks{This work is supported by NSFC (Grant Nos. 61272057,  61170270, 61309029),  Beijing Higher Education Young Elite Teacher Project (Grant Nos. YETP0475, YETP0477), BUPT Excellent Ph.D. Students Foundation (Grant No. CX201441).}
\thanks{The authors are with State Key Laboratory of Networking and Switching
Technology, Beijing University of Posts and Telecommunications,
Beijing 100876, China. Wei Huang is also with Science and Technology on Communication Security Laboratory, Chengdu, 610041, China. (e-mail: huangwei096505@yahoo.cn; wqy@bupt.edu.cn; lbhitmanbl@gmail.com; gaofei\_bupt@hotmail.com)

}}

\pagestyle{headings}

\maketitle

\begin{abstract}
A multi-user quantum key distribution protocol is proposed with single particles and the collective eavesdropping detection strategy on a star network. By utilizing this protocol, any two users of the network can accomplish quantum key distribution with the help of a serving center. Due to the utilization of collective eavesdropping detection strategy, the users of the protocol just need have the ability of performing certain unitary operations. Furthermore, we present three fault-tolerant versions of the proposed protocol, which can combat with the errors over different collective-noise channels. The security of all the proposed protocols is guaranteed by the theorems on quantum operation discrimination.
\end{abstract}


\IEEEpeerreviewmaketitle

\section{Introduction}

Over the last two decades, the principles of quantum mechanics have been widely applied in the field of information, which has promoted rapid developments of quantum cryptography and quantum computation. Since the  pioneering  work of  Bennett and Brassard in 1984 \cite{1_BB84}, quantum cryptography  has attracted a great deal of attention and has become one of the most promising applications of quantum information processing. There are several remarkable branches of quantum cryptography, including quantum key distribution (QKD) \cite{1_BB84,2_E91,3_92,4_TMTTT09,5_I06ie,6_Y09ie,7_-G.G.P,8_HK,9_PJW,10_WY,11_WXB,12_TM,13_HZF,14_ZGH.P}, quantum secure direct communication (QSDC) \cite{15_DFG,16_BF02,17_WC05,18-Cai03,19_DFG03,20_LXH06,21_CGL99}, quantum secret sharing (QSS) \cite{22_HBB99,23_KKI99,24_yang11,25_DZL05pla,26_ZKJ,27_ZZJ}, and secure multi-party computation (SMC) \cite{28_ZKJ06,29_QWMZ09,30_QPQ11,31-ZZ2000,32_Z2009}. As one of the most important parts of quantum cryptographic protocols, the multi-party quantum cryptographic protocol (MQCP) which involves at least three participants, such as quantum private comparison (QPC) protocols and QSS protocols, are more complicated than the  two-party ones. Therefore, more attention is needed in the research of MQCPs.

In most MQCPs, the quantum information carriers need to be transmitted for more than one time, and  usually the eavesdropping detection should be taken in every step of the transmission of them. However, this detection strategy, which is called step-by-step detection, always makes the protocols inefficient and complicated.  First, it is known that the security analysis of quantum cryptographic protocols is based on the error rate analysis with the theories in statistics. Hence, the proportion of the detection states (i.e., states chosen for eavesdropping) in the transmitted states should not be too small. If the detection is taken in every step of the transmission of the quantum information carriers, a lot of states will be used for checking eavesdropping and the qubit efficiency of the corresponding protocol will decrease with the increase of the number of detections. Second, detecting in every step of the transmission usually requires all the participants in such protocol to be equipped with many quantum devices, e.g., the quantum state measurement machine, the quantum state generation machine or the quantum storage machine. However, based on the current technology, these quantum devices are still expensive because of the difficulties on their constructions. As a consequence, it is uneconomical to require that every participant be equipped with most of these quantum devices. Apparently, MQCPs would be more efficient and easier to realize if the detection is taken only once in the whole procedure of the protocol. Fortunately, if a MQCP make use of collective eavesdropping detection strategy, it could meet such requirement. Collective detection is an efficient and useful eavesdropping detection strategy for MQCPs. On one hand, in a MQCP which employs the collective detection strategy, the detection needs to be taken only once after the whole transmission procedure of the quantum information carriers.  On the other hand, this strategy can also reduces the hardware requirement for the implementation of the protocol since all the users (except for the center who is responsible for preparing and measuring states) only need to perform certain unitary operations in the whole executing procedure of the protocol. To date, much attention have been focused on collective detection strategy and a lot of MQCPs haven been proposed by utilizing it (For simplicity, we will call the MQCP which uses collective detection MQCP-CD later) \cite{33_shil09,34_gao11,35_LB11,36_LB12,37_LB11,38_WTY11,39_LB, 40_HW12}.

In a MQCP-CD, all the users (except for the center) just need be capable of performing certain unitary operations. Therefore, the operations performed by them are very important to the security of the protocol. In this paper, a method for constructing the operations which are needed in the MQCP-CD is presented. It is a method that can be used to construct the unitary operations, which meet the security requirements of the MQCP-CD, with different kinds of quantum states, such as single photons, EPR pairs and  GHZ states. Based on this method,  we present a multi-user quantum key distribution (MQKD) protocol with collective detection and single particles. Our protocol is presented on a star network where any two of the involved users can execute quantum key distribution with the help of a serving center. There are several merits of this protocol. First, to establish a random key by employing this protocol, two users only need be capable of performing certain unitary operations. Second, none of the participants (all the users and the center) in our protocol need be equipped with the quantum storage machine. As storing quantum qubits is still a very difficult task in reality, our protocol is more feasible than the ones \cite{33_shil09,34_gao11,35_LB11} where the quantum storage machine is required. In addition, our protocol can resist various kinds of attacks from both the outside eavesdroppers and a dishonest center.

Actually, the quantum states transmitted  in channel will interact with the environment uncontrollably, which will introduce noise into the communication and influence both the correctness and efficiency of communication. If the variation of the noise is slower than the time delay between the quantum states transmitted  inside a time window, the states will be affected by the same noise. This kind of noise is called collective noise \cite{41_Bennett,42_Zanardi}. To combat with the errors caused by the collective noise, we further introduce three fault-tolerant versions of our protocol with the ideas of decoherence-free subspace (DFS) \cite{43_LinS14,44_LXH07,45_Duan,46_Boileau,47_HW,48_LXH08,49_Cabello,50_Sun Y,51_DFG.04,52_Walton,53_pgk00}, which can resist collective-dephasing noise, collective-rotation noise and all kinds of unitary collective noise, respectively.

The remainder of this paper is organized as follows. The next section presents our method for constructing the required unitary operations which can be used in designing MQCP-CD.  In sect. 3, our  MQKD  protocol and its three fault-tolerant versions, which  utilize collective detection and block transmission,  are proposed in detail. Block transmission, which was proposed firstly by Long et al. in \cite{15_DFG}, is one of the most important techniques for transmitting quantum states in quantum information processing. In block transmission, the quantum states are ordered and transmitted in blocks, and the eavesdropping detection is also executed on the blocks.  In sect. 4, the security of our proposed protocols is analyzed by using the theorems on quantum operation discrimination. Finally, a discussion as well as a short conclusion is given in Sect. 5.

\section{The method for constructing the unitary operations required in the MQCP-CD}

Thus far, many MQCP-CDs \cite{33_shil09,34_gao11,56_QSJ06,57_ZZJ,58_DFG} have been attacked since the unitary operations used in these protocols can be discriminated unambiguously (by a single use) if an eavesdropper utilizes some special attack strategies, such as dense-coding attack \cite{34_gao11} and fake-signal attack \cite{56_QSJ06}. It is just because there is still no effective method for constructing the required unitary operations that some improper ones were used in these protocols. In this section, we introduce a method for constructing the required unitary operations which can be used in designing a secure MQCP-CD in detail.  This method can be used to construct the required unitary operations with different kinds of quantum states. Afterwards, we prove the correctness of this method based on the conclusions on quantum operation discrimination.

\subsection{The detailed method}

Before presenting the method for constructing the required operations, we first introduce the basic principle of the MQCP-CD in brief \cite{33_shil09,34_gao11,35_LB11,36_LB12,37_LB11,38_WTY11,39_LB, 40_HW12}. In this kind of protocols, two mutually unbiased bases, which are denoted as $\{|a\rangle, |b\rangle\}$ and $\{|c\rangle, |d\rangle\}$, are required for secure communication. Here, $\langle a|b\rangle$=$\langle c|d\rangle$$=0$, $|\langle a|c\rangle|^{2}$=$|\langle a|d\rangle|^{2}$=$|\langle b|c\rangle|^{2}$=$|\langle b|d\rangle|^{2}$$=$$1/2$. Besides, there should be a center who is responsible for generating and measuring quantum states. The center first generates a sequence of states in the two bases and sends them to the first user. Then the first user processes the received states by performing four unitary operations according to his secret binary string and controlling binary string. After his/her operations, the first user sends the processed sequence to next user. Then the following users execute the procedure just like what the first user does one by one. When the last user finishes his/her operations, he/she sends the sequence back to the center. After the center receives the sequence, they  randomly choose some states to check eavesdropping with the information of the unitary operations performed on the chosen states and the corresponding measurement outcomes provided by the center. If the whole transmitting procedure is secure, the remaining states (or measurement outcomes) can be used to realize the main function of the protocol.

Concretely, when a user receives the sequence of quantum states, he/she first encodes his/her secret string by performing the operation $I$ (identity operator)/$U$ (encoding operation) on each of the received states if the corresponding bit of the secret string is 0/1. The effect of the unitary operations $U$ is that it flips a state in the same MB, i.e. $U|a\rangle=\alpha|b\rangle$, $U|b\rangle=\beta|a\rangle$, $U|c\rangle=\gamma|d\rangle$ and $U|d\rangle=\delta|c\rangle$. Here $\alpha$, $\beta$, $\gamma$ and $\delta$ are global phase factors which can be ignored. After that, he/she disturbs the encoded states by performing the operation $I$/$C$ (controlling operations) on each of these states  if the corresponding bit in his/her controlling string is 0/1. The effect of the unitary operations $C$ is that it flips each one of the four states in $\{|a\rangle$, $|b\rangle$, $|c\rangle$, $|d\rangle\}$ from one basis to the other basis.  When a user encodes his/her secret string on the received states, each of the bits in this secret string and controlling string will be used only once. If an eavesdropper wants to get some information of a bit in the secret string without leaving a trace in the eavesdropping detection, she/he should have the ability of discriminating the four unitary operations $I$, $U$, $C$ and $UC$ unambiguously with only one opportunity (i.e., under the condition that the device can be accessed only once). Therefore, one of the key steps in designing a secure MQCP-CD is to find appropriate unitary operations $U$ and $C$ which make $I$, $U$, $C$ and $UC$ impossible to be discriminated unambiguously with a single use.

Now we give the  method for constructing the required unitary operations. Suppose $V$ is a $d$-dimensional Hilbert space. By employing the Gram-Schmidt procedure, it is easy to construct an orthonormal basis of $V$, \{$|0'\rangle$, ..., $|(d-1)'\rangle$\}. It can be easily proved  that $\{|0'\rangle, |1'\rangle\}$ and $\{|+'\rangle, |-'\rangle\}$ are two mutually unbiased bases of a 2-dimentional subspace, where
\begin{eqnarray}
|+'\rangle=\frac{1}{\sqrt{2}}(|0'\rangle-i|1'\rangle), \quad|-'\rangle=\frac{1}{\sqrt{2}}(|1'\rangle-i|0'\rangle).
\end{eqnarray}
Then the encoding operation is chosen in the form of
\begin{eqnarray}
U=|0'\rangle\langle1'|+|1'\rangle\langle0'|+M,
\end{eqnarray}
where the part $M$ should meet the following two conditions. First, $M$ should be in a proper form in order to make $U$ be an unitary operation, i.e. $U^\dagger U$=$UU^\dagger$=$I$; Second, $M$ should be orthogonal to both $|0'\rangle$ and $|1'\rangle$. It can be easily verified that the operation $U$ could flips each one of the states in \{$|0'\rangle$, $|1'\rangle$, $|+'\rangle$, $|-'\rangle$\} in its own basis when $M$ is in a required form. There are many feasible choices for the form of $M$, such as $M=|2'\rangle\langle2'|+\cdots+|(d-1)'\rangle\langle (d-1)'|$ and $M=|2'\rangle\langle3'|+|3'\rangle\langle4'|+\cdots+|(d-1)'\rangle\langle2'|$. After getting the encoding operation $U$, we choose one of the square roots of $U$ as the controlling operation, which could flip each one of the states in \{$|0'\rangle$, $|1'\rangle$, $|+'\rangle$, $|-'\rangle$\} from one basis to the other basis, i.e. $C=$$\sqrt{U}$. The selection method for the operation $C$ will be given in the following proof. Thus far, a method for constructing the required encoding operation  and controlling operation has been introduced. That is to say, if the operations $U$ and $C$ are constructed by this method, the four unitary operations $I$, $U$, $C$ and $UC$ cannot be discriminated unambiguously with only one opportunity.

\subsection{Proof of the proposed method}

\begin{figure}
  \centering
\includegraphics[width=7cm]{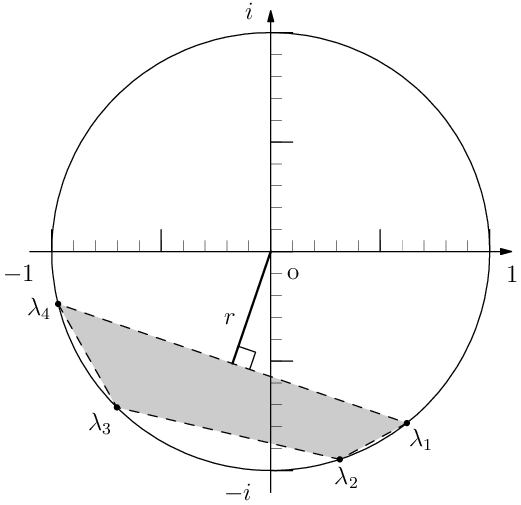}
  \caption{Take two-qubit unitary operations, for example. This figure  illustrates the definition of the function $r(U)=r$, where $\lambda_1$, $\lambda_2$, $\lambda_3$ and $\lambda_4$ are eigenvalues of the matrix $U$ and $r$ is the distance between the origin of the complex plane $o$ and the polygon $\lambda_1\lambda_2\lambda_3\lambda_4$. Obviously, $r=0$ indicates that $o$ is in/on the polygon $\lambda_1\lambda_2\lambda_3\lambda_4$..}\label{Fig.1}
\end{figure}

Herein we demonstrate that the four unitary operations $I$, $U$, $C$ and $UC$  cannot be discriminated unambiguously with only one opportunity if $U$ and $C$ are constructed by our method. Before giving the proof, we first introduce an important theorem on quantum operation discrimination.

\textbf{Theorem 1} \cite{60_mauro01} Under the condition that the device can be accessed only once, the minimum error probability to discriminate the two unitary operations $U_1$ and $U_2$ is
\begin{eqnarray}
P_e=\frac{1}{2}\left[1-\sqrt{1-4p_1p_2{\textrm{r}(U^\dagger_1U_2)}^2}\ \right],
\end{eqnarray}
where $\textrm{r}(U^\dagger_1U_2)$ stands for the distance between the origin of the complex plane and the polygon whose vertices are the eigenvalues of the unitary operator $U^\dagger_1U_2$ (see also Fig. 1), and $U^\dagger$ is the adjoint matrix of $U$.

\textbf{Corollary 2} Under the condition that the device can be accessed only once, two unitary operations $U_1$ and $U_2$ can be discriminated unambiguously if and only if $\textrm{r}(U^\dagger_1U_2)=0$.

As defined in the presented method, $C$=$\sqrt{U}$ ($U$=$C^2$). It can be easily found that $\textrm{r}(I^\dagger C)$=$\textrm{r}(U^\dagger UC)$=$\textrm{r}(C)$, $\textrm{r}(U^\dagger C)$=$\textrm{r}(U^{-1}C)$=$\textrm{r}(C^{-1})$=$\textrm{r}(C^\dagger)$. According to Theorem 1 and Corollary 2, under the condition that the device can be accessed only once, the two operations $U$ and $C$ ($I$ and $C$, $U$ and $UC$) constructed by our method cannot be discriminated unambiguously if and only if  neither of $\textrm{r}(C^\dagger)$ and $\textrm{r}(C)$ is equal to zero. Now we demonstrate that both $\textrm{r}(C^\dagger)$ and $\textrm{r}(C)$ are not equal to zero, i.e., $\textrm{r}(C^\dagger)$$>$$0$ and $\textrm{r}(C)$$>$$0$. Since $U$ is an unitary operation, all the eigenvalues of $U$ are the points on the unit circle in complex plane. Namely, all the eigenvalues of $U$ can be written in the form of $e^{i(\theta+2k\pi)}$, here $e^{i(\theta+2k\pi)}$=$e^{i\theta}$, $\theta$$\in$$[0,2\pi)$, $k$$\in$$Z$. Therefore, all the eigenvalues of the operation $\sqrt{U}$ should be in the form of $e^{i(\beta+k\pi)}$, here $\beta$=$\frac{\theta}{2}$$\in$$[0,\pi)$, $k$$\in$$Z$. It is obviously that $U$ has more than one square roots. Take the Pauli operation $\sigma_0$ as an simple example, since $\sigma_0$ can be written as either $e^{i\cdot0}|0\rangle\langle0|+e^{i\cdot0}|1\rangle\langle1|$ or $e^{i\cdot0}|0\rangle\langle0|+e^{i\cdot2\pi}|1\rangle\langle1|$, both $e^{i\cdot0}|0\rangle\langle0|+e^{i\cdot0}|1\rangle\langle1|$ and $e^{i\cdot0}|0\rangle\langle0|+e^{i\cdot\pi}|1\rangle\langle1|$ are the square roots of $\sigma_0$, i.e., $\sigma_0$=$\sigma_0^2$=$\sigma_z^2$. Here $|0\rangle$ and $|1\rangle$ represent the horizontal and vertical polarizations of photons, respectively.

As is shown above, all the eigenvalues of the operation $\sqrt{U}$ should be in the form of $e^{i(\beta+k\pi)}$, $k$$\in$$Z$. In the above method, we choose the  square root (of $U$) whose eigenvalues are all in the form of $e^{i\beta}$ (which means that the corresponding parameter $k$ is even) as the controlling operation $C$. As $\beta$$\in$$[0,\pi)$, all the the eigenvalues of $C$ are the points on the upper half of the unit circle (except for -1) in complex plane and therefore $\textrm{r}(C)$$>$$0$. In addition, since all the the eigenvalues of $C$ are the points on the upper half of the unit circle (except for -1), it is evident that all the the eigenvalues of $C^\dagger$ are the points on the bottom half of the unit circle (except for -1), which means $\textrm{r}(C^\dagger)$$>$$0$. That is to say, the two operations $U$ and $C$ ($I$ and $C$, $U$ and $UC$) constructed by our method cannot be discriminated unambiguously under the condition that the device can be accessed only once.

Till now, we have proved that the four unitary operations $I$, $U$, $C$ and $UC$ constructed by the above method cannot be discriminated unambiguously under the condition that the device can be accessed only once. Utilizing this method, we can construct the required unitary operations for designing a secure MQCP-CD with different kinds of quantum states.

\subsection{The role of the proposed method}

If someone wants to design a MQCP-CD using a certain kind of quantum state in his/her favour, one of the most important things he/she should do first is to find the corresponding encoding operation ($U$) and controlling operation ($C$), which satisfy the security requirement of the MQCP-CD. Obviously, the method we just proposed above can be used to construct such unitary operations according to the quantum states that the protocol designer wants to use. For example, if he/she wants to design a MQCP-CD with single photons, a natural choice for one of the two bases could be \{$|0\rangle$, $|1\rangle$\}. By employing our method, the other basis could be chosen as$\{|+\rangle$, $|-\rangle\}$, where $|+\rangle=\frac{1}{\sqrt{2}}(|0\rangle-i|1\rangle)$ and $|-\rangle=\frac{1}{\sqrt{2}}(|1\rangle-i|0\rangle)$, and the corresponding encoding operation $U_s$ and controlling operation $C_s$ can be constructed  as
\begin{eqnarray}
U_{s}=|0\rangle\langle1|+|1\rangle\langle0|
      =\left(
           \begin{array}{cc}
             0 & 1 \\
             1 & 0 \\
           \end{array}
         \right),\nonumber\\
C_{s}=\sqrt{U_{s}}=\frac{1+i}{2}\left(
                      \begin{array}{cc}
                        1 & -i \\
                        -i & 1 \\
                      \end{array}
                    \right).
\end{eqnarray}
It can be easily verified that \{$|0\rangle$, $|1\rangle$\} and $\{|+\rangle, |-\rangle\}$ form two mutually unbiased bases. The effect of the operations $U_s$ and $C_s$ on the four states can be illustrated as
\begin{eqnarray}
&U_s|0\rangle=|1\rangle, \quad U_s|1\rangle=|0\rangle,\nonumber\\
&U_s|+\rangle=|-\rangle, \quad U_s|-\rangle=|+\rangle, \nonumber\\
&C_s|0\rangle=(\frac{1+i}{\sqrt{2}})|+\rangle, \quad C_s|1\rangle=(\frac{1+i}{\sqrt{2}})|-\rangle, \nonumber\\
&C_s|+\rangle=(\frac{1-i}{\sqrt{2}})|1\rangle, \quad C_s|-\rangle=(\frac{1-i}{\sqrt{2}})|0\rangle.
\end{eqnarray}

Utilizing the two bases and two operations $(U_{s}$ and $C_{s})$ given above, one can design different kinds of MQCP-CDs (QSS, QPC, etc.) with single photons. Evidently, there are still some other choices for the two operations if we choose other single particles to be the quantum information carriers. To date, in order to improve the qubit efficiency or reduce the hardware requirement of the users, some MQCP-CDs \cite{33_shil09,34_gao11,35_LB11,36_LB12,37_LB11,38_WTY11,39_LB,40_HW12} have been proposed with single photons. Unfortunately, all of these protocols need to store quantum qubits, which is still a difficult task in reality. To make use of collective detection under the current techniques, we present a more feasible MQKD protocol without employing quantum storage machine in next section. More importantly, we also enhance the proposed protocol to be immune to the errors over different collective-noise channel based on the above method.

\section{The proposed MQKD protocols}

\vspace*{4mm}
\begin{figure}
  \centering
\includegraphics[width=7cm]{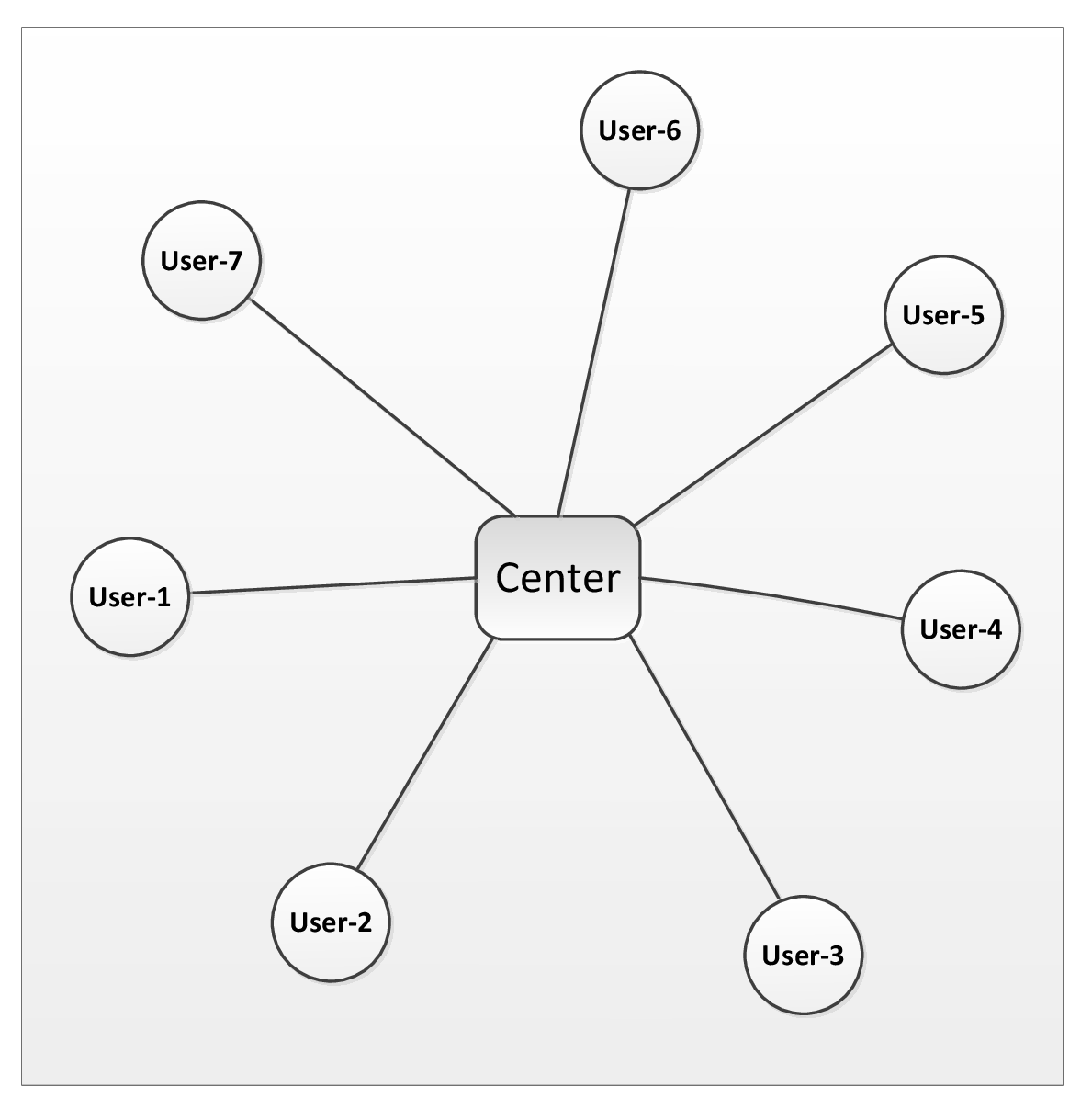}
  \caption{A simple illustration of the MQKD with 7 users on a star network. In this network, any two of the 7 users can establish a shared secret key only with unitary operations.}\label{Fig.2}
\end{figure}
\vspace*{4mm}

In this section, we present a MQKD protocol by employing single particles and collective detection on a star network. In this protocol, there is a center who is responsible for generating and measuring quantum states. With the help of the center, any two users involved in the network can securely establish a random key just by performing unitary operations on the states transmitted to them.  If user-$i$ wants to share a random key with user-$j$, user-$i$ and user-$j$ can encode their random binary strings into the states produced by the center, and then they are able to deduce a random key with the measurement outcomes published by the center, where 1$\leq$$i$, $j$$\leq$$n$. In this circumstance, user-$i$ and user-$j$ only need to hide their secrets in the transmitted states with proper unitary operations. In this network, we assume that any two of the participants (the center and all the users) are able to transmit quantum states between each other. Similar to most of the previous quantum cryptographic protocols, the classical channels involved in this protocol are supposed to be authenticated.  Compared with the existing MQKD protocols which also utilize collective detection \cite{33_shil09,34_gao11,35_LB11}, our protocol have the following two advantages. On one hand, it can be secure against the attacks from both the outside eavesdroppers and a dishonest center. On the other hand, it need not make use of quantum storage machine, which indicates that our protocol is more feasible in practice under the current techniques.

\subsection{The proposed MQKD protocol with single particles}

\begin{figure}
  \centering
\includegraphics[width=9cm]{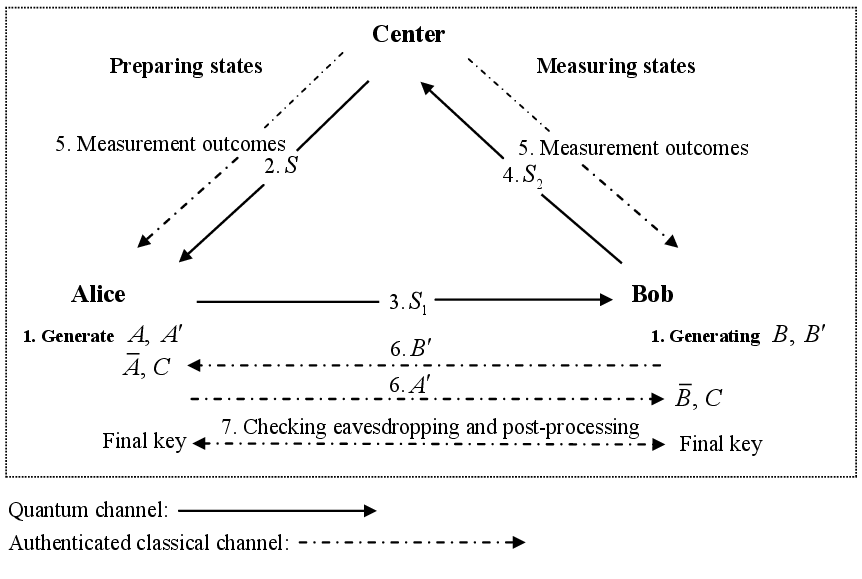}
  \caption{The subsystem of the presented QKD network with single particles.}\label{Fig.3}
\end{figure}

\vspace*{4mm}

\begin{center}
\tabcolsep=20pt  
\small
\renewcommand\arraystretch{1.2}  
\begin{minipage}{8.5cm}{
\small{\bf Table 4.} The relationship among the values of $\bar{A_j}$, $\bar{B_j}$, $\bar{A_j^\prime}$$+$$\bar{B_j^\prime}$ and $C_j$ when no errors occur.}
\end{minipage}
\vglue5pt
\begin{tabular}{| c | c | c | c | }  
\hline  
 {$\bar{A_j}$} & {$\bar{B_j}$} & {$\bar{A_j^\prime}$$+$$\bar{B_j^\prime}$} & {$C_j$} \\     
\hline
  {$\bar{A_j}$} & {$\bar{B_j}$} & {0 or 1} & {$\bar{A_j}$$\oplus$$\bar{B_j}$}  \\   
  \hline
  {$\bar{A_j}$} & {$\bar{B_j}$} & {2} & {$\bar{A_j}$$\oplus$$\bar{B_j}$$\oplus$1}  \\   
\hline
\end{tabular}
\end{center}

Assuming that two users involved in the network, say Alice and Bob, want to share a random key, they could execute the MQKD protocol  described as below (see also Fig. 3).

\begin{itemize}
\item[1:]Alice generates two random binary strings of length $4n$, which are denoted as $A$ and $A^\prime$, respectively. Similarly, Bob generates two random binary strings of length $4n$, which are denoted as $B$ and $B^\prime$, respectively. After that, Alice informs the center that she wants to establish a random key with Bob.
\item[2:]After receives Alice's request, the center prepares a sequence of $4n$ single particles which are all in the state $|0\rangle$ (denoted as sequence $S$) and sends it to Alice.
\item[3:] Once receiving $S$,  Alice performs the unitary operations $U^{A_i}$ and $C^{A_i^\prime}$ on the $i$-th particle in $S$, for 1$\leq$$i$$\leq$$4n$. Here, $A_i$ and $A_i^\prime$  are respectively the $i$-$th$ bit in strings $A$ and $A^\prime$. $U^1$=$U_s$, $C^1$=$C_s$ and $U^0$=$C^0$=$I_s$ is the identity density operator on two-dimensional Hilbert space, i.e., $I_s$=$|0\rangle\langle0|$+$|1\rangle\langle1|$. After that,  she sends the new sequence (denoted as $S_1$) to Bob.
\item[4:]Upon receiving the sequence $S_1$, Bob  performs the operations $U^{B_i}$ and $C^{B_i^\prime}$ on the $i$-th particle in $S_1$, for 1$\leq$$i$$\leq$$4n$.  Then he sends the the new sequence (denoted as $S_2$) back to the center.
\item[5:]Once receiving $S_2$, the center makes measurement on each of the particles randomly in $\sigma_z$-basis or $\sigma_y$-basis, where $\sigma_z$-basis=\{$|0\rangle$,$|1\rangle$\} and $\sigma_y$-basis=\{$|+\rangle$,$|-\rangle$\}. Afterwards, he/she publishes the measurement outcome of each of the particles in $S_2$. According to the $i$-th measurement outcome announced by the center, Alice and Bob could learn which of the two bases was used to measure the $i$-th particle in $S_2$. Concretely, if the measurement outcome is $|0\rangle$ or $|1\rangle$ ($|+\rangle$ or $|-\rangle$), $\sigma_z$-basis  ($\sigma_y$-basis) was used.
\item[6:]After the center announced the measurement outcomes of all the particles in $S_2$, Alice and Bob publish $A^\prime$ and $B^\prime$, respectively, where $A_i^\prime$ ($B_i^\prime$) indicates that whether Alice (Bob) has performed $C_s$ on the $i$-th particle in the travelling sequence. Based on the announced information, Alice and Bob are able to determine which of the particles in $S_2$ were measured in correct bases. Here measuring the $i$-th particle in correct basis represents that the $i$-th measurement outcome is $|0\rangle$ or $|1\rangle$ ($|+\rangle$ or $|-\rangle$) under the condition that  $A_i^\prime$+$B_i^\prime$ is 0 or 2 (1). According to probabilistic theory, half of the particles in $S_2$ (i.e., 2 $n$ particles in $S_2$) have been measured by the center with correct measuring bases. For the positions of the measurement outcomes obtained from incorrect measuring bases,  Alice and Bob discard the corresponding bits in $A$, $B$, $A^\prime$ and $B^\prime$, where the new strings are denoted as $\bar{A}$, $\bar{B}$, $\bar{A^\prime}$ and $\bar{B^\prime}$, respectively. Then Alice and Bob could deduce a 2$n$-bit string $C$ with the measurement outcomes obtained with correct measuring bases. Concretely, if the measurement outcome is $|0\rangle$ or $|+\rangle$ ($|1\rangle$ or $|-\rangle$), the corresponding bit of $C$ is 0 (1). The relationship among the values of $\bar{A_j}$, $\bar{B_j}$, $\bar{A_j^\prime}$$+$$\bar{B_j^\prime}$ and $C_j$ when no error occurs is shown in Table 4, where 1$\leq$$j$$\leq$2$n$ and $\oplus$ denotes the addition module 2.
 \item[7:]To check eavesdropping, Bob randomly chooses $n$ positions out of string $C$ and requires Alice to tell him the corresponding bits in $\bar{A}$. According to the information announced by Alice and Table 4, Bob checks whether the corresponding bits in $C$ are in accordance with the theoretical values. If more than an acceptable error rate is found, they abort the results; otherwise, Bob can trust the transmission and deduce the rest $n$ bits of $\bar{A}$ with the corresponding bits of $C$ and $\bar{B}$. At last, Alice and Bob utilize error correction and privacy amplification \cite{54_Gisin N,55_Inamori H} to establish the secure session key.
\end{itemize}

In this protocol, with the help of the center, any two of the involved users can establish a random key only with unitary operations. Although the qubit efficiency of our protocol is half as those of the protocols in Refs. \cite{33_shil09,34_gao11,35_LB11}, our protocol need not store quantum qubits. Hence, our protocol is more feasible with the the current techniques. Besides, the users should set up the filter and the beam splitter to prevent the Trojan horse attack and invisible-photon attack \cite{63_Cai06,64-LXH06}.

\section{Fault-tolerant MQKD protocols against collective noise}

Herein we introduce three fault-tolerant versions of the proposed MQKD protocol based on the method in Sect. 2 and the idea of DFS, which can be immune to the collective-dephasing noise, collective-rotation noise and all kinds of unitary collective noise, respectively.

\subsection{Fault-tolerant MQKD protocol against collective-dephasing noise}

The collective-dephasing noise \cite{41_Bennett,44_LXH07,65-ZZJ14} can be described as
\begin{eqnarray}
      |0\rangle\longrightarrow|0\rangle,\quad\quad\quad |1\rangle\longrightarrow e^{i\phi}|1\rangle,
 \end{eqnarray}
where $\phi$ is the noise parameter and it fluctuates with time. A logical qubit, which is composed with two physical qubits with antiparallel parity (as bellow),  is immune to collective-dephasing noise as both the logical qubits obtains the same phase factor $e^{i\phi}$ through this kind of channel.
\begin{eqnarray}
      |0\rangle_L=|0\rangle|1\rangle,\quad\quad\quad |1\rangle_L=|1\rangle|0\rangle.
 \end{eqnarray}
To communicate securely, at least two non-orthogonal measuring bases are required. According to the presented method, one basis could be \{$|0\rangle_L$, $|0\rangle_L$\}, and the other one is \{$|+\rangle_L$, $|-\rangle_L$\}, where
\begin{eqnarray}
      |+\rangle_L=\frac{1}{\sqrt{2}}(|0\rangle_L-i|1\rangle_L),\quad|-\rangle_L=\frac{1}{\sqrt{2}}(|1\rangle_L-i|0\rangle_L).
 \end{eqnarray}
It can be easily verified that $|\langle+|0\rangle_L|^{2}$=$|\langle+|1\rangle_L|^{2}$=$|\langle-|0\rangle_L|^{2}$=\\$|\langle-|1\rangle_L|^{2}$=$\frac{1}{2}$, which indicates that $\{|0\rangle_L$, $|1\rangle_L\}$ and $\{|+\rangle_L$, $|-\rangle_L\}$ form two mutually unbiased bases.
Based on the method given in Sect. 2, we construct the encoding operation $U_{dp}$ and controlling operation $C_{dp}$ for our MQKD protocol, which can resist collective-dephasing noise, as follows:
\begin{eqnarray}
U_{dp}&=&|0\rangle_{LL}\langle1|+|1\rangle_{LL}\langle0|+M\nonumber\\
      &=&|01\rangle\langle10|+|10\rangle\langle01|+|00\rangle\langle11|+|11\rangle\langle00|\nonumber\\
      &=&\left(
      \begin{array}{cccc}
           0 & 0 & 0 & 1 \\
           0 & 0 & 1 & 0 \\
           0 & 1 & 0 & 0 \\
           1 & 0 & 0 & 0
         \end{array}
         \right),\nonumber\\
C_{dp}&=&\sqrt{U_{dp}}=\frac{1+i}{2}\left(
  \begin{array}{cccc}
    1 & 0 & 0 & -i \\
    0 & 1 & -i & 0 \\
    0 & -i & 1 & 0 \\
    -i & 0 & 0 & 1
  \end{array}
\right).
\end{eqnarray}
The effect of the operations $U_{dp}$ and $C_{dp}$ on the four states can be illustrated as
\begin{eqnarray}
&U_{dp}|0\rangle_L=|1\rangle_L,\quad U_{dp}|1\rangle_L=|0\rangle_L,\nonumber\\
&U_{dp}|+\rangle_L=|-\rangle_L,\quad U_{dp}|-\rangle_L=|+\rangle_L,\nonumber\\
&C_{dp}|0\rangle_L=(\frac{1+i}{\sqrt{2}})|+\rangle_L,\quad C_{dp}|+\rangle_L=(\frac{1-i}{\sqrt{2}})|1\rangle_L,\nonumber\\
&C_{dp}|1\rangle_L=(\frac{1+i}{\sqrt{2}})|-\rangle_L,\quad C_{dp}|-\rangle_L=(\frac{1-i}{\sqrt{2}})|0\rangle_L.
\end{eqnarray}

As in this case with collective-dephasing noise, with the help of the center, any two of the involved users can establish a shared secret key over collective-dephasing channel with the same steps of the protocol given in Sect. 3.1. Certainly, there are some differences between these two cases. On one hand,  the four states $|0\rangle$, $|1\rangle$, $|+\rangle$ and $|-\rangle$ should be respectively replaced by $|0\rangle_L$, $|1\rangle_L$, $|+\rangle_L$ and $|-\rangle_L$. On the other hand, the unitary operations $I_s$, $U_s$ and $C_s$ should be substituted with $I_s^{\otimes2}$, $U_{dp}$ and $C_{dp}$, respectively, where $I_s^{\otimes2}$=$I_s$$\otimes$$I_s$. Accordingly, the measuring bases used by the center in step 5 should be replaced by \{$|0\rangle_L$, $|1\rangle_L$\} and \{$|+\rangle_L$, $|-\rangle_L$\}.

\subsection{Fault-tolerant MQKD protocol against collective-rotation noise}

The collective-rotation noise \cite{41_Bennett,44_LXH07,65-ZZJ14} can be described as
\begin{eqnarray}
      &&|0\rangle\longrightarrow\cos\theta|0\rangle+\sin\theta|1\rangle,\nonumber\\
     &&|1\rangle\longrightarrow-\sin\theta|0\rangle+\cos\theta|1\rangle,
 \end{eqnarray}
where $\theta$ is the parameter of noise which fluctuates with time.  The two Bell states, $|\Phi^+\rangle$=$\frac{1}{\sqrt{2}}$$(|00\rangle$$
+$$|11\rangle)$ and $|\Psi^-\rangle$=$\frac{1}{\sqrt{2}}(|01\rangle$$-$$|10\rangle)$, are invariant under this collective-rotation noise. Naturally, logical
qubits under this noise can be chosen as
\begin{eqnarray}
      |0_{r}\rangle_L=|\Phi^+\rangle,\quad\quad\quad |1_{r}\rangle_L=|\Psi^-\rangle.
 \end{eqnarray}
For secure communication, at least two non-orthogonal measuring bases  are required. According to the presented method, one basis could be \{$|0_r\rangle_L$, $|0_r\rangle_L$\}, and the other one is \{$|+_r\rangle_L$, $|-_r\rangle_L$\}, where
\begin{eqnarray}
      &&|+_{r}\rangle_L=\frac{1}{\sqrt{2}}(|0_{r}\rangle_L-i|1_{r}\rangle_L),\nonumber\\
      &&|-_{r}\rangle_L=\frac{1}{\sqrt{2}}(|1_{r}\rangle_L-i|0_{r}\rangle_L).
 \end{eqnarray}
It is easy to verify that $|\langle_{r}+|0_{r}\rangle_L|^{2}$=$|\langle_{r}+|1_{r}\rangle_L|^{2}$=$|\langle_{r}-|0_{r}\rangle_L|^{2}$\\=$|\langle_{r}-|1_{r}\rangle_L|^{2}$=$\frac{1}{2}$, \, which indicates that $\{|0_{r}\rangle_L$, $|1_{r}\rangle_L\}$ and $\{|+_{r}\rangle_L$, $|-_{r}\rangle_L\}$ form two mutually unbiased bases.
Using the method given in Sect. 2, we construct the encoding operation $U_r$ and controlling operation $C_r$ for the  MQKD protocol over collective-rotation channel as follows:
\begin{eqnarray}
U_{r}&=&|0_r\rangle_{LL}\langle1_r|+|1_r\rangle_{LL}\langle_r0|+M_r\nonumber\\
      &=&|\phi^+\rangle\langle\psi^-|+|\psi^-\rangle\langle\phi^+|+|\psi^+\rangle\langle\phi^-|+|\phi^-\rangle\langle\psi^+|\nonumber\\
      &=&\left(
      \begin{array}{cccc}
       0 & 1 & 0 & 0 \\
       1 & 0 & 0 & 0 \\
       0 & 0 & 0 & -1 \\
       0 & 0 & -1 & 0
         \end{array}
         \right),\nonumber\\
C_{r}&=&\sqrt{U_r}=\frac{1+i}{2}\left(
  \begin{array}{cccc}
     1 & -i & 0 & 0 \\
     -i & 1 & 0 & 0 \\
     0 & 0 & 1 & i \\
     0 & 0 & i & 1
  \end{array}
\right).
\end{eqnarray}
The effect of the operations $U_r$ and $C_r$ on the four states can be illustrated as
\begin{eqnarray}
&U_r|0_{r}\rangle_L=|1_{r}\rangle_L,\quad U_r|1_{r}\rangle_L=|0_{r}\rangle_L,\nonumber\\
&U_r|+_{r}\rangle_L=|-_{r}\rangle_L,\quad U_r|-_{r}\rangle_L=|+_{r}\rangle_L,\nonumber\\
&C_r|0_{r}\rangle_L=(\frac{1+i}{\sqrt{2}})|+_{r}\rangle_L,\quad C_r|+_{r}\rangle_L=(\frac{1-i}{\sqrt{2}})|1_{r}\rangle_L,\nonumber\\
&C_r|1_{r}\rangle_L=(\frac{1+i}{\sqrt{2}})|-_{r}\rangle_L,\quad C_r|-_{r}\rangle_L=(\frac{1-i}{\sqrt{2}})|0_{r}\rangle_L.
\end{eqnarray}

As in this case with collective-rotation noise, with the help of the center, any two of the involved users can establish a shared secret key over collective-rotation channel with the same steps of the protocol presented in Sect. 3.1. Of course, there are some differences between these two cases. One difference is that the four states $|0\rangle$, $|1\rangle$, $|+\rangle$ and $|-\rangle$ should be respectively replaced by $|0_r\rangle_L$, $|1_r\rangle_L$, $|+_r\rangle_L$ and $|-_r\rangle_L$. Another difference is that the unitary operations $I_s$, $U_s$ and $C_s$ should be substituted with $I_s^{\otimes2}$, $U_r$ and $C_r$, respectively. Accordingly, the measuring basis used by the center in step 5 should be replaced by basis \{$|0_r\rangle_L$, $|1_r\rangle_L$\} and \{$|+_r\rangle_L$, $|-_r\rangle_L$\}.

\subsection{Fault-tolerant MQKD protocol against all kinds of unitary collective noise}

Decoherence-free (DF) states \cite{42_Zanardi,53_pgk00} is a type of states which are changeless under any $n$-lateral unitary transformation (i.e., $U^{\otimes n}|\psi^-\rangle=|\psi^-\rangle$, where $U^{\otimes n}$=$U$$\otimes$$\cdots$$\otimes$$U$ denotes the tensor product of $n$ unitary transformations $U$). The amount of quantum information that a given $N$-qubit DFS is able to protect depends on the number of its qubits. For $N$ even, the DFS spanned by states which are eigenstates of the whole Hamiltonian of the qubit-bath system has dimension \cite{42_Zanardi,49_Cabello}
\begin{eqnarray}
d(N)=\frac{N!}{(N/2)!(N/2+1)!}
\end{eqnarray}
For $N$=2, there exists only one DF state, the singlet state $|\psi^-\rangle$. For $N$=4, the dimension of the DFS is 2. Hence, 4 qubits are sufficient to fully protect one arbitrary logical qubit from all kinds of unitary collective noise \cite{49_Cabello}. A natural choice for the orthogonal basis of the 4-qubit DFS is
\begin{eqnarray}
|\bar{0}\rangle_L&=&|\psi^-\rangle_{12}|\psi^-\rangle_{34}\nonumber\\
&=&\frac{1}{2}(|0101\rangle+|1010\rangle-|0110\rangle-|1001\rangle)_{1234},\nonumber\\
|\bar{1}\rangle_L&=&\frac{1}{2\sqrt{3}}(2|0011\rangle+2|1100\rangle-|0101\rangle\nonumber\\
&&-|1010\rangle-|0110\rangle-|1001\rangle)_{1234}.
\end{eqnarray}
To communicate securely, at least two non-orthogonal measuring bases are required.  According to the presented method, One basis could be \{$|\bar{0}\rangle_L$, $|\bar{1}\rangle_L$\}, and the other should be chosen as \{$|\bar{+}\rangle_L$, $|\bar{-}\rangle_L$\}, where
\begin{eqnarray}
      |\bar{+}\rangle_L=\frac{1}{\sqrt{2}}(|\bar{0}\rangle_L-i|\bar{1}\rangle_L),\quad
      |\bar{-}\rangle_L=\frac{1}{\sqrt{2}}(|\bar{1}\rangle_L-i|\bar{0}\rangle_L).
 \end{eqnarray}
It can be easily verified that $|\langle\bar{+}|\bar{0}\rangle_L|^{2}$=$|\langle\bar{+}|\bar{1}\rangle_L|^{2}$=$|\langle\bar{-}|\bar{0}\rangle_L|^{2}$
=$|\langle\bar{-}|\bar{1}\rangle_L|^{2}$=$\frac{1}{2}$, which means $\{|\bar{0}\rangle_L,$ $|\bar{1}\rangle_L\}$ and $\{|\bar{+}\rangle_L, |\bar{-}\rangle_L\}$ form two mutually unbiased bases. Let us suppose $W$ is the 4-qubit Hilbert space whose dimension is 16, then we are able to find an orthonormal basis \{$|\bar{0}\rangle_L$, $|\bar{1}\rangle_L$, $|\bar{15}\rangle$\} for $W$ by employing the Gram-Schmidt procedure.  For the sake of simplicity, we do not give the concrete form of the states $|\bar{2}\rangle_L$, $|\bar{3}\rangle_L$, ..., $|\bar{15}\rangle$ as the well known Gram-Schmidt procedure is not complicated. Once obtaining all the states of the orthonormal basis, we can construct the corresponding encoding operation $\bar{U}$ and controlling operation $\bar{C}$ for our robust MQKD protocol based the presented method. Specifically, the encoding operation $\bar{U}$ and controlling operation $\bar{C}$ are in the form as
\begin{eqnarray}
 \bar{U}=|\bar{0}\rangle\langle\bar{1}|+|\bar{1}\rangle\langle\bar{0}|+O, \quad\quad \bar{C}=\sqrt{\bar{U}}.
 \end{eqnarray}
Here, we have many feasible choices for the from of $O$, such as $O=|\bar{2}\rangle\langle\bar{2}|+,...+|\bar{15}\rangle\langle \bar{15}|$ and $O=|\bar{2}\rangle\langle\bar{3}|+|\bar{3}\rangle\langle\bar{4}|+...+|\bar{15}\rangle\langle\bar{2}|$.
For example, when $O=|\bar{2}\rangle\langle\bar{3}|+|\bar{3}\rangle\langle\bar{4}|+...+|\bar{15}\rangle\langle\bar{2}|$, the effect of the operations $\bar{U}$ and $\bar{C}$ on the states can be illustrated as below.
\begin{eqnarray}
&\bar{U}|\bar{0}\rangle_L=|\bar{1}\rangle_L,\quad \bar{U}|\bar{1}\rangle_L=|\bar{0}\rangle_L, \nonumber\\
&\bar{U}|\bar{+}\rangle_L=|\bar{-}\rangle_L,\quad \bar{U}|\bar{-}\rangle_L=|\bar{+}\rangle_L,\nonumber\\
&\bar{C}|\bar{0}\rangle_L=(\frac{1+i}{\sqrt{2}})|\bar{+}\rangle_L,\quad \bar{C}|\bar{+}\rangle_L=(\frac{1-i}{\sqrt{2}})|\bar{1}\rangle_L,\nonumber\\
&\bar{C}|\bar{1}\rangle_L=(\frac{1+i}{\sqrt{2}})|\bar{-}\rangle_L,\quad \bar{C}|\bar{-}\rangle_L=(\frac{1-i}{\sqrt{2}})|\bar{0}\rangle_L.
\end{eqnarray}

As in this case with all kinds of unitary collective noise, with the help of the center, any two of the involved users can establish a shared secret key over all kinds of collective-noise channels with the same steps described in the protocol protocol proposed in Sect. 3.1. Certainly, there are some differences between these two cases. That is, the four states $|0\rangle$, $|1\rangle$, $|+\rangle$ and $|-\rangle$ should be respectively substituted with $|\bar{0}\rangle_L$, $|\bar{1}\rangle_L$, $|\bar{+}\rangle_L$ and $|\bar{-}\rangle_L$, and the unitary operations $I_s$, $U_{dp}$ and $C_{dp}$ should be replaced by $I_s^{\otimes4}$, $\bar{U}$ and $\bar{C}$,  respectively. Accordingly, the center should utilize the bases \{$|\bar{0}\rangle_L$, $|\bar{1}\rangle_L$\} and \{$|\bar{+}\rangle_L$, $|\bar{-}\rangle_L$\} to measure each of the states in $S_2$ in step 5.

\section{Security analysis}
\label{sec:3}

In this section, we analyze the security of the MQKD protocol with single particles. For the security of the ones over collective-noise channels, it can be analyzed in the same way with the case with single particles. For clarity, we first consider the attacks from outside eavesdroppers. After that, we take into account the situation where a center tries to eavesdrop the key.

\subsection{Security against outside attacks }
\noindent
Let us assume that Eve is an attacker who wants to eavesdrop the users'  secret key without being noticed in the eavesdropping detection. Eve could intercept the traveling particles sent to the receiver and replace them with the ones prepared by herself \cite{56_QSJ06}, or she can entangle the travelling particles with additional states and try to extract information from these states \cite{66-gf10}. Since the secret strings of the users are encoded in the operations performed on the travelling particles, the action to eavesdrop the a user's secret strings is equivalent to discriminate the operations that he/she has performed. For instance, if Eve wants to get the value of $A_i$ (1$\leq$$i$$\leq$4$n$), she should know which one of the operations $I_s$, $U_s$, $C_s$ and $U_s$$C_s$ Alice has performed on the corresponding particle. In other words, Eve should be capable of discriminating the four unitary operations.  Actually, conclusions on quantum operation discrimination have been well studied. In addition to Theorem 1 and Corollary 2, here we introduce another conclusion.

\textbf{Theorem 3} \cite{61_Wang2006} The quantum operations $\gamma_1$, ..., $\gamma_n$ can be unambiguously discriminated by a single use if and only if for
any $i$=1, 2, ..., $n$, supp($\gamma_i$) $\nsubseteq$ supp($S_i$), where supp($\gamma$) denotes the support of a quantum operation $\gamma$ and $S_i=\{\gamma_j:j\neq i\}$.

In the proposed protocol with single particles, the unitary operations performed by both the two users (Alice and Bob) can be viewed
as four unitary operations as a whole, i.e. $I_s$, $U_s$, $C_s$ and $U_sC_s$. Here, $I_s$ is the identity density operator on two-dimensional Hilbert space and the operations $U_s$ and $C_s$ are respectively the encoding operation and controlling operation defined in equation (4). It could be easily found that these operations satisfy the following relationship, $C_s$=$\frac{1+i}{2}$$(1\cdot I_s-i\cdot U_s$+$0\cdot U_sC_s)$$=\frac{1+i}{2}(I_s-iU_s)$, which indicates that supp$\{C_s\}$ $\subseteq$ supp$\{I_s$, $U_s$, $U_sC_s\}$. Therefore, these four operations cannot be unambiguously discriminated by a single use (i.e., with only one opportunity) according to Theorem 3. Moreover, we can also get the same conclusion according to Theorem 1. For example, the eigenvalues of the operation $U_{s}^\dag C_{s}$ are $1$ and $-i$, therefore, ${r(U_{s}^\dag C_{s})=1/\sqrt{2}}$ and the minimum error probability to discriminate $U_{s}$ and $C_{s}$ is
\begin{eqnarray}
P_e=\frac{1}{2}\left[1-\sqrt{1-{(1/\sqrt{2})}^2}\ \right]\approx0.15
\end{eqnarray}
Just in this way, we can also get that the minimum error probability to discriminate $I_s$ and $C_{s}$/$I_s$ and $U_{s}C_{s}$/$U_s$ and $U_{s}C_{s}$ is $P_e$$(\approx0.15)$, which means that these operations cannot be discriminated unambiguously under the condition that the device can be accessed only once.

In the proposed protocol, the users' secret strings are encoded in the operations they performed on the states. If Eve wants to get some information of a bit of a user's secret string without leaving a trace in the eavesdropping detection, she should have the ability of unambiguously finding out which of  $I_s$, $U_s$, $C_s$ and $U_sC_s$ has been performed on the corresponding particle with only one opportunity. However, as we analyzed above, the four operations cannot be unambiguously discriminated under this condition.  Consequently, the well-known attacks, such as intercept-resend attack, measurement-resend attack, entanglement-measure attack and dense-coding attack,  from an outside eavesdropper will inevitably be detected in the eavesdropping detection.

Furthermore, as for the two special attacks for two-way communication, i.e., Trojan horse attack and invisible-photon attack, the users and center can makes use of the methods in \cite{63_Cai06,64-LXH06} to protect the proposed protocols from such attacks. Hence,  we omit the redundant description here.

\subsection{Security against the center's attacks}

It is known that the attacks from of a dishonest participant is more powerful than those from the outside eavesdroppers. On one hand, he/she knows part of the legal information. On the other hand, he/she could tell lies in the the executing procedure of the protocol in order to avoid introducing errors into the eavesdropping detection. Therefore, the attacks from dishonest participants should be paid more attention to. In fact, such situations should also be considered in our protocol. First, in the case that the enter has been corrupted by others, it may try to eavesdrop the random key shared between the users. Second,  in some special situations, the center may be out of service and the person who has the ability of generating and measuring the quantum states may want to eavesdrop the key. That is, the one who is able to substitute the center may be dishonest. Now, we show that, if the users executing the proposed protocol, a dishonest center cannot get the random key without leaving a trace in the eavesdropping detection,

Compared with the outside eavesdroppers, a dishonest center has the following two superiorities. First, he/she could replace the particles in $S$ with whatever kind of states he/she wants. Second, he/she could modify the measurement outcomes which are announced  in step 5. Nevertheless, he/she is still unable to get the key shared between by the users without being noticed in the eavesdropping detection of our protocol. According to the analysis given above, no matter what kind of states the dishonest center prepared in step 2, he/she is unable to unambiguously discriminate the four unitary operations that performed by the users under the condition that the device can be accessed only once. In other words, these four unitary operations cannot be precisely discriminated when they are performed on a single qubit or one qubit of any entangled state, respectively. That is to say, if the center utilizes a strategy to  to discriminate these four operations, he/she will get incorrect outcome with a non-negligible probability. In our protocol,  the users will announce the controlling string $A^\prime$ and $B^\prime$ only after the center publishes the measurement outcomes of all the particles in $S_2$. Also the bits utilized to check eavesdropping is randomly chosen by the users after the measurement outcomes being published. Since the dishonest center cannot precisely discriminate the four unitary operations utilized in our protocol with a single use, no matter what attack  strategy he/she employs, once he/she could get part of the useful information about the secret string, he/she will inevitably introduce errors into the eavesdropping detection and hence be noticed by the users.

Till now, we have analyzed the security of the MQKD protocol in sect. 3.1 to show that it is secure against attacks from both the outside eavesdroppers and the dishonest center. Of course, as for the cases of collective-dephasing noise/collective-rotation noise/all kinds of unitary collective noise, we can also show that it could be immune to all the present attacks just in the same way, since the operations $I^{\otimes2}$, $U_{dp}$, $C_{dp}$ and $U_{dp}C_{dp}$ / $I^{\otimes2}$, $U_r$, $C_r$ and $U_rC_r$/$I^{\otimes4}$, $\bar{U}$, $\bar{C}$ and $\bar{U}\bar{C}$ cannot be discriminated unambiguously (under the condition that the device can be accessed only once) according to both the theorems given above. Thus, we do not elaborate on the proof processes for simplicity.

\section{Discussion and conclusion}
\noindent

\subsection{Discussion}

To combat with the errors over collective-noise channel, Zhang has proposed a well-known fault-tolerant multiparty quantum secret sharing (MQSS) protocol \cite{27_ZZJ}. It is easy to find that this MQSS protocol can also be used for MQKD. However, there are clear differences between this protocol and our protocol. First, by utilizing our protocol, two arbitrary users could establish a secret key with the help of a center, and then they could utilize this key for secure communication between each other. In this circumstance, the center is unable to get any useful information of the shared key. While by employing Zhang's protocol, a boss could establish a joint key with his/her agent entirety. Only when all the agents collaborate can they deduce the joint key that has been shared with the boss and then utilize this key for secure communication with the boss. Second, as there exists a serving center in our protocol, if two of the involved users  want to establish a secret key, they only need be capable of performing certain unitary operations. Nevertheless, in Zhang's protocol, if the boss wants to establish a joint key with his/her agent entirety, the boss should be able to generate quantum states, and the agent entirely should be capable of performing a certain unitary operation and measuring quantum states. In addition, to establish a secret key, the quantum states sequence should be transmitted three times in our protocol, while the sequence in Zhang's protocol only need be transmitted twice.

In fact, with some minor modifications, the protocol proposed in Sect. 3 can also be used for secret sharing of a random key. Concretely, if we want this protocol to be used for QSS, the following modifications are needed. Firstly, in step 5, after the center finish measuring all the received particles, he/she notifies the fact to Alice and Bob. Different to the original protocol, he/she no longer publishes the measurement outcomes. Secondly, in step 6, once receiving the center's notification, Alice and Bob publish $\bar{A^\prime}$ and $\bar{B^\prime}$. With this information, the center judges which of the received particles have been measured in correct measuring basis. Then he/her informs Alice and Bob of the positions of the particles which have been measured incorrectly. For the positions of the measurement outcomes obtained from incorrect measuring bases, Alice and Bob discard the corresponding bits in $A$, $B$, $A^\prime$ and $B^\prime$, and obtain the new strings which are denoted as $\bar{A}$, $\bar{B}$, $\bar{A^\prime}$ and $\bar{B^\prime}$, respectively. Also, the center deduces a 2$n$-bit string  $C$ according to the corresponding  2$n$ measurement outcomes obtained from correct measuring bases. The relationship among the bit values of $\bar{A_j}$, $\bar{B_j}$, $\bar{A_j^\prime}$$+$$\bar{B_j^\prime}$ and $C_j$  is shown in Table 4, where 1$\leq$$j$$\leq$2$n$. Thirdly, in step 7, the center randomly chooses $n$  positions out of string $C$ and requires Alice and Bob to tell him/her the corresponding bits in  $\bar{A}$ and $\bar{B}$, respectively. According to the received information and Table 4, the center checks whether there exist eavesdropping in the executing procedure of the protocol. If there exists no eavesdropping, the center has successfully established a joint key   with Alice and Bob. That is to say, only when Alice and Bob collaborate can they first establish a joint key with center and then extract the secret messages from the enter's encrypted messages later transmitted via a public channel.

So far, we have shown that the protocol proposed in Sect. 3 could be used for three-party QSS with some minor modifications. Nevertheless, the security requirement for QSS has certain differences with QKD. On one hand, the boss (i.e., the message sender) of a QSS protocol is honest. On the other hand, the agents (i.e., sharers) of a QSS protocol may be dishonest. That is to say, some dishonest agents may cooperate to attack the protocol and try to get the key without the help of other agents. According to the security analysis given in Sect. 4, it is not hard to find that the three-party QSS we just mentioned above is secure. However, when the number of the agents is more than 2, many more threatening attacking strategies for QSS, such as entanglement swapping attack \cite{38_WTY11}, should be considered. That is to say, to extend the above three-party protocol to an $n$-party one, some extra strategies should be employed for resisting these attacks. As the related strategies have been discussed somewhere \cite{39_LB}, we do not focus on this issue here.

\subsection{Conclusion}

In this paper, we introduce a method for constructing the encoding operations and controlling operations, which are required in the MQCP-CD. Then by employ single particles and collective detection, we present a MQKD protocol on a star network without storing qubits, which can resist the attacks from both the outside eavesdroppers and the dishonest center. Based on the proposed method and the idea of DFS, we also introduce three fault-tolerant versions of the proposed protocol over collective-dephasing noise channel, collective-rotation noise channel and all kinds of unitary collective noised channel.

Obviously, the method we presented in this paper is useful as it can be  used to construct the unitary operations required in the MQCP-CD with different kinds of quantum states. Compared with the the existing MQCP-CDs \cite{33_shil09,34_gao11,35_LB11,36_LB12,37_LB11,38_WTY11,39_LB,40_HW12}, the  protocols proposed in this paper have the following advantages. First, the proposed protocols are more feasible since they do not need to employ quantum storage machine. Second, the proposed protocols can not only utilize collective detection, but also be immune to the collective noise.

\ifCLASSOPTIONcaptionsoff
  \newpage
\fi

\end{document}